\documentclass{elsart5p}

 \usepackage{graphicx}

\usepackage{amssymb}

\def\beq{\begin{equation}}
\def\eeq{\end{equation}}
\def\bea{\begin{eqnarray}}
\def\eea{\end{eqnarray}}
\def\vd{\langle v_d q \rangle}

\newcommand*{\eqref}[1]{Eq.~(\ref{eq:#1})}
\newcommand*{\eqlab}[1]{\label{eq:#1}}
\newcommand*{\figref}[1]{Fig.~\ref{fig:#1}}
\newcommand*{\figlab}[1]{\label{fig:#1}}

\newcommand*{\secref}[1]{Section~\ref{sec:#1}}
\newcommand*{\seclab}[1]{\label{sec:#1}}

\def\VYP#1#2#3{{\bf #1}, #3 (#2)}  

\def\PL#1#2#3{Phys.~Lett.~\VYP{#1}{#2}{#3}}

\begin{document}

\begin{frontmatter}



\title{Macroscopic Model of Geomagnetic-Radiation from Air Showers}


\author[KVI]{Olaf Scholten} and
\ead{scholten@kvi.nl}
\author[SUBA]{Klaus Werner}
\address[KVI]{Kernfysisch Versneller Instituut, University
of Groningen, 9747 AA, Groningen, The Netherlands}
\address[SUBA]{SUBATECH,
University of Nantes -- IN2P3/CNRS-- EMN,  Nantes, France}

\begin{abstract}
We have developed a macroscopic description of coherent electro-magnetic
radiation from air showers initiated by ultra-high energy cosmic rays in the
presence of the geo-magnetic field. This description offers a simple and direct
insight in the relation between the properties of the air shower and the
time-structure of the radio pulse. As we show the structure of the pulse is a
direct reflection of the important length scales in the shower.
\end{abstract}

\begin{keyword}
Radio detection \sep Air showers \sep Cosmic rays \sep Geo-synchrotron \sep
Geo-magnetic \sep Coherent radio emission

\PACS 95.30.Gv \sep 95.55.Vj \sep 95.85.Ry \sep 96.50.S- \sep
\end{keyword}
\end{frontmatter}


\section{Introduction}

In recent years the interest in the use of radio detection for cosmic ray air
showers is increasing with the promising results obtained from recent
LOPES~\cite{Fal05,Ape06} and CODALEMA~\cite{Ard06} experiments, which triggered
plans to install an extensive array of radio detectors at the Pierre Auger
Observatory~\cite{Ber07}. For these reasons there is a strong interest to
understand the link between the properties of the extensive air shower (EAS) and
the time structure of the emitted pulse.

Already in the earliest works on radio emission from air
showers~\cite{Jel65,Por65,Kah66,All71}, the importance of coherent emission was
stressed. Two mechanisms, Cherenkov radiation and geo-magnetic radiation were
investigated. In more recent work~\cite{Fal03,Sup03,Hue05}, the picture of
coherent synchrotron radiation from secondary electrons and positrons in the
Earth's magnetic field was proposed and extensive results on geo-synchrotron
emission are given in~\cite{Hue07}.

The emission of an electromagnetic pulse by the charges in an EAS is described by
Classical Electrodynamics~\cite{Jac-CE}. In this sense it is an easy matter. The
complication arises is the designation of all possible charge and current
distributions that drive the electromagnetic emission process. For this models
are needed such as the collective model described in \secref{MGMR}. This
macroscopic description of geomagnetic radiation~\cite{Sch08,Wer08} (MGMR) is in
the basic idea very similar to that used in Ref.~\cite{Kah66}. However,
independent of any particular model, some of the general features of the emitted
pulse can be calculated based the length scales involved in an EAS as is
discussed in \secref{L-scales}. For simplicity of the discussion we will limit
ourselves here to an ideal case of a vertical airshower (along the $z$-axis)
while the Earth-magnetic field is taken parallel to the Earth (along the
$y$-axis). The case for more realistic geometry can be found in the
literature~\cite{Wer08}.

\section{A game of length scales\seclab{L-scales}}

Our primary interest lies in the {\emph coherent} emission of radio waves from an
EAS. In a coherent process the amplitudes of the individual emitting charges
interfere constructively such that the intensity of the radiation, the square of
the amplitude, is proportional to the quare of the number of charges, $N^2$.
Since the typical number of charges in an EAS is of the order of $N=10^6$ it is
clear that this process is far more important than incoherent emission where the
intensity is just proportional to $N$.

Classical Electromagnetism~\cite{Jac-CE} teaches that the coherent emission of a
system of charges decreases linearly for wave-length which are considerably
longer than the the size of the emitting body. For wave-length that are
considerably shorter than that of the charge distribution the emission spectrum
is truncated. The latter is due to destructive interference of radiation emitted
from charges that are more than a wavelength apart. The length scales of the
source thus put strong limitation on the frequency spectrum of the emitted
radiation.

An EAS, independent how enormous the event may be, always exists for a limited
time and occurs in a limited part of the atmosphere. The atmosphere it is assumed
to be charge neutral free from electrical currents before the cosmic ray entered.
The same applies to the atmosphere some time after the EAS has touched the
ground. All electro dynamics thus occurs in a limited region of space-time and
this has as an immediate consequence that the response of the system, the emitted
radio waves, should vanish linearly in the limit of infinite wave length or zero
frequency. Thus the time integral of the positive and negative parts of the
emitted pulse should be equal. The simplest structure of the pulse is thus
bi-polar but -in principle- it can have more than a single zero crossing.

The high frequency cut-off of the coherent response is at wave length that are
shorter than the largest length scale of interest. Some of the length scales that
may be important in this respect are i) the thickness of the front of the EAS,
the pancake thickness, ii) the projected length of the EAS in the direction of
the observer, iii) the lateral distribution of the charges in the EAS, and as
last iv) something different from the previous three. Which length scale is
determining the pulse shape is one of the challenges for models of radio emission
from EAS. The high-frequency cut-off is reflected the short-time structure of the
emitted pulse, the time between the start and the zero crossing.

\section{The Macroscopic Model for Geo-Magnetic Radiation\seclab{MGMR}}

The front of an EAS is formed by a plasma with copious amounts of electrons,
positrons and other particles all moving towards the surface of the Earth with a
velocity $c \beta_s$, almost the light velocity ($\beta_s\approx 1$). The
magnetic field of the Earth induces, through the Lorentz force which pulls the
electrons and positrons in opposite directions, a net electric current in the
electron-positron plasma. This mechanism is very similar to that inducing an
electric current in a copper wire that is moved through a magnetic field.  The
applied force induces a constant drift velocity due to collisions of the charge
carriers with the air molecules (EAS) or copper atoms (wire), where the value
depends on the strength of the magnetic field. Please notice that there are also
large differences between electrons in metals, where the average velocity is due
to thermal motion, and electrons in an EAS, where main component of the velocity
is non-thermal an is attenuated with shower age. This macroscopic geomagnetic
radiation model (MGMR) is discussed in detail in Ref.~\cite{Sch08,Wer08}, here
only the main findings are reproduced.

For the present estimate it is assumed that there are equal numbers of positive
and negative charges moving towards the Earth with a large velocity. The number
of electrons and positrons in the shower is parameterized as $N(z)=N_e f_t(t_r)$
in terms of the normalized shower profile, $f_t(t_r)$, at height $-z=c \beta_s
t_r$ and where $N_e$ is the number of electrons in the shower at the maximum. Due
to the Earth's magnetic field a net electrical current in the $\hat{x}$-direction
is induced with magnitude
\beq
 j(x,y,z,t)= \vd \, e\, N_e f_t(t_r) \;,
\eqlab{CurrDens}
\eeq
where the pancake thickness is ignored. The drift velocity depends rather
strongly on the model assumptions made~\cite{Sch08,Wer08} and we adopt a value of
$\vd=0.04$~c.

From the current density, $j^{\mu}$, the vector potential is given by the
Li\'{e}nard-Wiechert fields,
\begin{equation}
 A^{\mu}(x)= \frac{1}{4 \pi \varepsilon_0} \int
 \left. {j^\mu\over R(1-\vec{\beta}\cdot \hat{n})}\right|_{\mbox{ret}} \,dh\;,
 \eqlab{LW}
\end{equation}
for a source with an infinitesimally small lateral extension. We use the common
notation where $\hat{n}$ is a unit vector pointing from the source to the
observer and $R$ is the distance, both evaluated at the retarded time.

To obtain a simple estimate for the emitted radiation we take the limit
$\beta_s=1$ and $n=1$ and ignore the thickness of the pancake. In this limit the,
denominator in \eqref{LW} can be rewritten to give
\bea
{\cal D} &=& R(1-\vec{\beta_s}\cdot \hat{n})|_{\mbox{ret}}
 =  \sqrt{(-c\beta_s t +h)^2 + (1-\beta_s^2 n^2)d^2} \nonumber \\
 &=& c\beta_s t  + {\cal O}(1-\beta_s^2) \approx ct
 \eqlab{denom} \;.
 \eea
Positive values of $t$ correspond to negative retarded times,
\beq
ct_r={ct\over 1+\beta_s} - {d^2 \over 2 c\beta_s t}+ {\cal O}(1-\beta_s^2)
 \approx -{d^2 \over 2 c t}  \;,
 \eqlab{t-ret-app}
\eeq
where $t=0$ corresponds to the time the EAS touches Earth. From this equation we
obtain the interesting observation that the earlier times of the received pulse
is emitted at large (and negative) retarded times and thus large heights while
the tail part of the pulse is emitted when the EAS was already close to the
ground.

Since the current density has only an $\hat{x}$-component, the vector potential
will share this property,
\beq
A^x(t,d)= J {f_t(t_r) \over {\cal D}} \;,\eqlab{Ax}
\eeq
where $J= \vd N_e e/ 4 \pi \varepsilon_0 c$. The electric and magnetic fields are
obtained from the vector potential by taking derivatives. Since in the present
example the zeroth component of the vector potential vanished (no excess charge)
the electric field is proportional to the time derivative of $\vec{A}$, giving
\beq E_x(t,d)
 \approx J {c^2 t_r^2 4 \over d^4} {d\over dt_r}[ t_r f_t(t_r)] \;.
\eqlab{E-appx}
\eeq
Since the vector potential vanishes at very early and (almost) vanishes at late
retarded times, it is clear that the time-integrated electric field (being the
derivative of $A$) also vanishes. This is in agreement with the observation made
in \secref{L-scales}.

One striking aspect of \eqref{E-appx} is that the pulse is given as a simple
function of $t_r$. Because of the relation \eqref{t-ret-app} the pulse at
different distances is the same function of $t_r=-d^2/(2c^2 t)$ i.e.\ at twice
the distance from the point of impact it is four times as wide while the
amplitude is decreased by a factor $2^4$, see curves labeled 'appx' in
\figref{Pulse-effects-t17}.

Another interesting aspect is the realization that the zero crossing of the
electric field corresponds to the maximum in the vector potential (in fact $t_r
f_t(t_r)$). This part of the pulse thus corresponds to radio emission from the
EAS at an height exceeding that of the shower maximum. The dominant part of the
pulse is thus emitted at heights well above the shower maximum which implies that
the radio signal has information on the early stages of the EAS development.

For small distances to the shower core, $d\approx ct$, the approximations made in
deriving the expression for the retarded time, \eqref{t-ret-app}, are no longer
valid and \eqref{E-appx} is thus not applicable. More elaborate
calculations~\cite{Sch08} show that for observers close to the shower core, the
pancake thickness become the important length scale, see blue curve in
\figref{Pulse-effects-t17}.

\begin{figure}[ht]
 \centerline{\includegraphics[height=6cm,bb=20 135 480 800,clip]{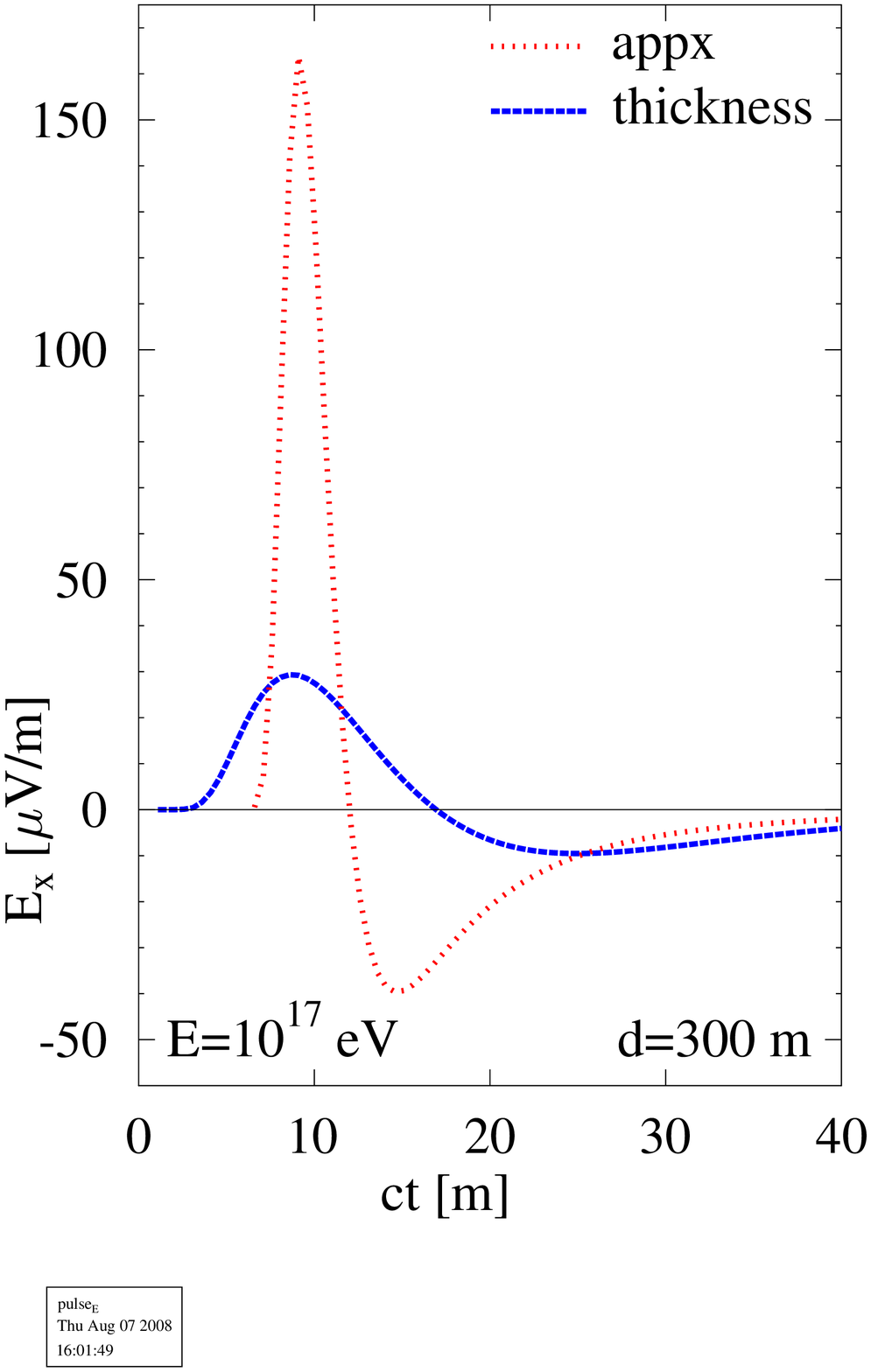}
 \
 \includegraphics[height=6cm,bb=20 135 480 800,clip]{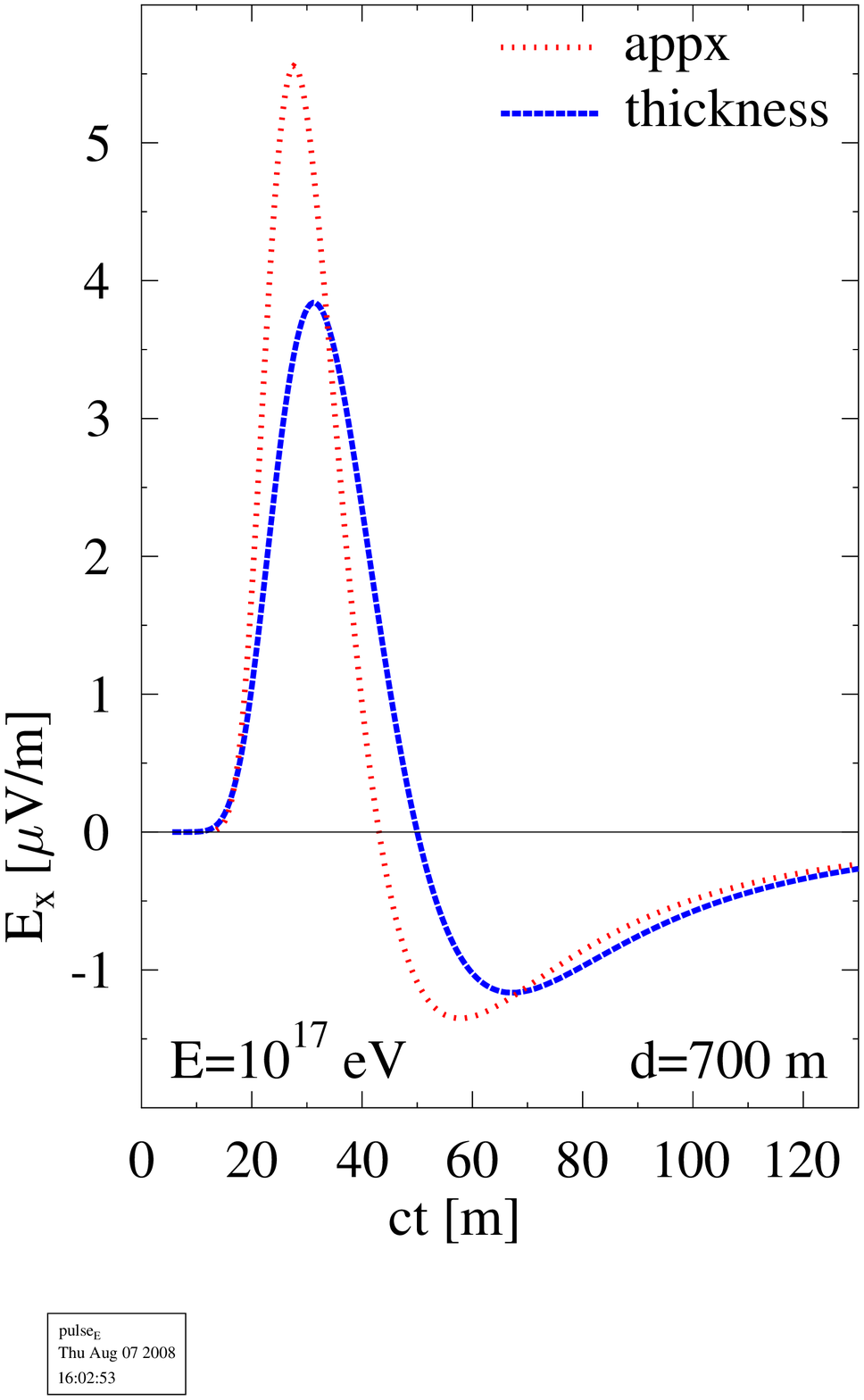}}
\caption[fig17-eff]{[color online]
\it Predicted pulse shapes at $300$~m and
$700$~m from the shower core for a $10^{17}$~eV shower. The dotted curve labeled 'appx'
corresponds to the limiting case of \eqref{E-appx} while the drawn curve
includes the effects of the finite pancake thickness~\cite{Sch08}.}
  \figlab{Pulse-effects-t17}
\end{figure}

\section{Conclusions}

We have shown that the macroscopic model for radio emission offers a very clear
and simple picture for the emitted radio pulse. Different length scales determine
the structure of the radio pulse, depending on the distance from the shower core.
Close to the core the pulse length is determined by the pancake thickness while
at larger distances the shower profile is reflected in the pulse structure. From
the radio pulse at various distances from the shower core thus both the pancake
thickness as well as the height of the shower maximum can be determined. This
offers interesting prospects to determine the cosmic-ray composition.

The present discussion has been limited to the main contribution to the emitted
signal coming from the electrical current induced by the Earth magnetic field. In
Ref.~\cite{Wer08} also the contribution are investigated which are due to induced
electric dipole moments and charge excess for realistic EAS simulations. These
are shown to give correction of the order of 10\%, leaving the basic conclusions
unchanged. Also the effect of more realistic geometries and a finite index of
refraction of air have been studied.


\begin{thebibliography}{00}





\bibitem{Fal05} H. Falcke, \etal,
Nature \VYP{435}{2005}{313}.

\bibitem{Ape06} W. D. Apel, \etal,
Astropart. Physics \VYP{26}{2006}{332}.

\bibitem{Ard06} D. Ardouin, \etal,
Astropart. Physics \VYP{26}{2006}{341}.

\bibitem{Ber07} A.M. van den Berg for the Pierre Auger Collaboration, ICRC 2007
    abstract

\bibitem{Jel65} J. V. Jelley \etal, Nature \VYP{205}{1965}{327}.

\bibitem{Por65} N.A. Porter, C.D. Long, B. McBreen, D.J.B Murnaghan and T.C.
    Weekes, \PL{19}{1965}{415}.

\bibitem{Kah66} F.D. Kahn and I.Lerche, Proc. Royal Soc. London
    \VYP{A289}{1966}{206}.

\bibitem{All71} H. R. Allan, 
Prog. in Element. part. and Cos. Ray Phys. \VYP{10}{1971}{171}.

\bibitem{Fal03} H.~Falcke and P.~Gorham, Astropart. Phys. \VYP{19}{2003}{477}.

\bibitem{Sup03} 
D.A. Suprun, P.W. Gorham, J.L. Rosner, Astropart. Phys. \VYP{20}{2003}{157}.

\bibitem{Hue05} T. Huege, H. Falcke, Astronomy \& Astrophysics
    \VYP{430}{2005}{779}; T. Huege, H. Falcke,  Astropart. Phys.
    \VYP{24}{2005}{116}.

\bibitem{Hue07} T. Huege, R. Ulrich, and R. Engel, 
    Astropart. Phys. \VYP{27}{2007}{392}

\bibitem{Jac-CE} J.D. Jackson, {\emph Classical Electrodynamics}, Wiley, New
    York, 1999.

\bibitem{Sch08} Olaf Scholten, Klaus Werner, and Febdian Rusydi, Astropart.\
    Phys.\ \VYP{29}{2008}{94}.

\bibitem{Wer08} Klaus Werner and Olaf Scholten, Astropart.Phys.\
    \VYP{29}{2008}{393}, arXiv:0712.2517 [astro-ph]

\end{thebibliography}
\end{document}